# Fractionally delta-doped oxide superlattices for higher carrier mobilities

*Woo Seok Choi*[1], *Suyoun Lee*[1,2], *Valentino R. Cooper*[1], *and Ho Nyung Lee*[1]

[1]Materials Science and Technology Division, Oak Ridge National Laboratory, Oak Ridge, TN 37831, USA

[2]Electronic Materials Research Center, Korea Institute of Science and Technology, Seoul 136-791, Korea

**A 2D electron gas system in an oxide heterostructure serves as an important playground for novel phenomena. Here, we show that, by using fractional $\delta$-doping to control the interface's composition in $La_xSr_{1-x}TiO_3/SrTiO_3$ artificial oxide superlattices, the filling-controlled 2D insulator-metal transition can be realized. The atomic-scale control of *d*-electron band filling, which in turn contributes to the tuning of effective mass and density of the charge carriers, is found to be a fascinating route to substantially enhanced carrier mobilities.**

Two dimensional electron gas (2DEG) is one of the essential ingredients in semiconductor technologies, which is utilized in various electronic devices including field effect transistors and charge coupled devices.[1] Recently, complex-oxide-based 2DEGs have shown the potential for discovering unexpected phenomena, which might lead to development of novel oxide-based information and energy technologies.[2,3] In particular, $SrTiO_3$- (STO-) based 2DEGs exhibit a wide spectrum of physical properties, including interface superconductivity, magnetism, electric field controlled resistance change, and large thermoelectric power, originating from strong many-body interactions.[3-5]

One of the particularly interesting STO-based 2DEGs is the $LaTiO_3/SrTiO_3$ (LTO/STO) heterostructure, which is composed of the $3d^0$ band insulator STO and the $3d^1$ Mott insulator LTO. Surprisingly, a 2D metallic state is induced near the interface when the two insulators are adjoined together, due to the smearing out of the $3d^1$ electron in LTO towards the $3d^0$ band of STO, so called the “electronic reconstruction”.[6-14] Similarly, $La_xSr_{1-x}TiO_3$ (LSTO) bulk solid solutions also exhibit a metallic state. The latter comes from the classic filling-controlled insulator-metal transition (IMT) by adding electrons with increasing $x$ to the empty $3d^0$ band of STO.[15,16] As $x$ approaches one, it becomes a Mott insulator.

In this letter, we report a combination of experiment and theory on the filling-controlled IMT in an atomically-designed 2D superlattice (SL). We fabricated high quality, fractionally delta- ($\delta$-) doped $[(La_xSr_{1-x}TiO_3)_1/(SrTiO_3)_{10}]_{10}$ (LSTO/STO) SLs, in order to tune the carrier density ($n$) and effective mass ($m^*$) of 2DEG as schematically shown in Figure 1a. In addition to the observation of the La/Sr ratio ($x$) dependent IMT, we could obtain enhanced carrier mobility ($\mu$) by selectively incorporating low mass bands. We found that $x = 0.50$ SL showed decreased $n$ and $m^*$ comparable to those of weakly interacting electrons in semiconductor 2DEGs, revealing the highest $\mu$ among the systems we have studied and similar LTO/STO integer SLs.[13] While $\mu$ itself cannot be compared to other record high values,[17,18] our work provides an important step towards further enhancing the electronic properties of oxide 2DEGs.

Fractional LSTO/STO SLs were fabricated using pulsed laser epitaxy. The state-of-the-art growth enabled us to precisely control beyond the layer-by-layer synthesis for fractionally $\delta$-doped SLs with sub-monolayer accuracy. More details on the experimental methods and theory are summarized in the Supporting Information. Figure 1b shows oscillations of the reflective high-energy electron diffraction (RHEED) specular spot intensity obtained during the fabrication of an $x$ = 0.5 SL. First, the yellow region indicates the growth of 10 unit cell (u.c.) thick STO, where a period of 10 u.c. is pronouncedly counted. At this point, the RHEED specular spot intensity is at a local maximum. Second, the green region indicates the growth of 0.5 u.c. thick LTO, where the ablation of LTO is ceased when the RHEED specular spot intensity reaches a local minimum. Third, the blue region indicates the growth of 0.5 u.c. thick STO. The RHEED specular spot intensity is counted until it recovers to the local maximum. (The second and third steps together build up one u.c. thick $La_{0.5}Sr_{0.5}TiO_3$.) This procedure is then repeated for ten times to grow a $[(La_{0.5}Sr_{0.5}TiO_3)_1/(STO)_{10}]_{10}$ SL. The final SLs had extremely flat surfaces with a well conserved ~0.4 nm high step-and-terrace structure, mimicking the surface of the $TiO_2$-terminated STO substrate, as shown in atomic force microscopy surface topography (inset of Figure 1b). Therefore, we believe that the fractional layers of LTO were grown in the form of random 2D nano-dots (one u.c. thick) consisting of perovskite atomic building blocks,[19] which were then covered by compensating fractional layers of STO as schematically shown in Figure 1a.

Figures 1c and d show x-ray diffraction (XRD) $\theta$-$2\theta$ patterns and a reciprocal space map (RSM) around the substrate's 002 and 114 peaks, respectively, for the LSTO/STO SLs. The pronounced peaks and well-defined Kiessig fringes in the XRD scans indicate the excellent crystallinity of the SLs with the designed-in periodicity as well as atomically smooth surfaces and sharp interfaces. The intensity of the SL peaks decreased systematically with $x$, since the chemical contrast between the layers became less distinct. The small shift of the $00\underline{22}$ peak to a higher angle with decreasing $x$ also suggests that our SLs have a systematic variation in the $c$-axis lattice constant as anticipated (LTO has a slightly larger lattice constant as compared to that of STO.). Moreover, the RSM indicates that the SLs are fully strained without any in-plane lattice relaxations and are of high quality, as the separation of XRD peaks from the Cu-$K\alpha_1$ and Cu-$K\alpha_2$ radiation is clearly present.

Figure 2a shows temperature ($T$) dependent sheet resistance ($R_{xx}(T)$). The $x = 0$ SL was highly insulating (not shown as the $R_{xx}$ is beyond the instrumental limit, *i.e.*, >2 MΩ), confirming that our samples were free from oxygen vacancies or cation off-stoichiometry, which might cause unwanted conducting carriers.[20] The $x = 0.25$ SL also revealed an insulating/semiconducting behavior with a high resistance. This insulating/semiconducting behavior might arise from the fact that the coverage of LTO was below the percolation threshold for a metallic layer. As $x$ increases further, an IMT occurred and the SL started to show a metallic $T$-dependent behavior. This indicates that for $x > \sim 0.25$ a sufficient portion of electrons in LTO smear out to neighboring STO layers, resulting in an overall metallic $\delta$-doped layer. As $x$ increases, $R_{xx}(T)$ decreased systematically, suggesting that $n$ and/or $\mu$ of the SL were increasing (See Supporting Information for more detailed discussions).

To obtain quantitative information on $n$ and $\mu$, we measured the magnetic field dependent Hall resistance ($R_{xy}(H)$) as shown in Figures 2b-e. For $x = 0.25$ SL, the $R_{xy}(H)$ curve showed a linear behavior up to 10 T. On the other hand, $x = 0.50$, 0.75, and 1.00 SLs showed a strong non-linear behavior below ~50 K. Such non-linear behavior in $R_{xy}(H)$ has not been observed in uniformly mixed LSTO solid solution films.[21] However, in low dimensional LTO/STO systems such as integer SLs, it has been successfully interpreted with the multi-channel conduction model.[13] For the two-channel transport case,

$$R_{xy}(H) = [(\mu_1^2 n_1 + \mu_2^2 n_2) + (\mu_1\mu_2 B)^2(n_1 + n_2)] \,/\, e[(\mu_1|n_1| + \mu_2|n_2|)^2 + (\mu_1\mu_2 B)^2(n_1 + n_2)^2]. \qquad (1)$$

Note that multi-channel conduction has also been considered for explaining the transport properties of other STO-based heterostructures.[4,12,22]

We used Eq. (1) to fit $R_{xy}(H)$ with the constraint of $R_{xx}(H = 0) = 1/e(n_1\mu_1 + n_2\mu_2)$ and interpreted the non-linear Hall behavior.[13] The fitting results (lines) in Figures 2b-e were in good agreement with the experimental data (symbols). Thus, we conclude that the transport behavior of the SLs indeed comes from the contribution of at least two different types of carriers, namely, high-density-low-mobility (HDLM) carriers (carrier 1) and low-density-high-mobility (LDHM) carriers (carrier 2). The resulting $n$ and $\mu$ are shown in Figures 3a and b, respectively. The carrier density of HDLM carriers ($n_1$) had weak $T$-dependence (except for $x$ = 0.50 SL, which will be discussed later in detail) where it slightly decreases with decreasing $T$. The LDHM carriers started to appear below ~50 K, where the non-linearity became evident in $R_{xy}(H)$, and its carrier density ($n_2$) increased with decreasing $T$. Above that $T$, $n_2$ was too small to be detected by our Hall measurements. The mobilities of both HDLM carriers ($\mu_1$) and LDHM carriers ($\mu_2$) increased with decreasing $T$ and then saturated below ~20 K.

Now we focus on the $x$-dependence of the LSTO/STO SLs. Figures 3c and d summarize the $x$-dependent $n$ and $\mu$, respectively, at 300 and 2 K. Note that Figure 3c is shown in $n_{2D}$ for better comparison, which corresponds to $n$ of 2DEG for a single LSTO layer, instead of ten 2DEGs from the entire SL shown in Figure 3a. While LDHM carriers did not show any discernible or systematic $x$-dependence, HDLM carriers showed systematic changes at 2 K as shown in Figure 3c: $n_1$ ($\mu_1$) increased (decreased) with increasing $x$ (We focus on the data at 2 K, where the extrinsic contribution from the thermal vibration, *i.e.* phonon, is minimized.). Interestingly, $n_1$ for the $x$ = 0.5 SL showed a substantial decrease at low $T$ compared to that in other SLs. The reduced $n_1$ in turn implies a drastic increase in $\mu_1$. Indeed, an order of magnitude enhancement in $\mu_1$ (reaching a value of 115 $cm^2$/Vs) was observed for the $x$ = 0.5 SL .

In STO-based heterostructures with the multichannel conduction, it has been found that the carriers with higher $\mu$ are located in the STO side away from the interface, due to the enhanced dielectric screening by high dielectric constant STO at low $T$. For example, in a $LaAlO_3$/STO system, high $\mu$ carriers are formed in STO, while majority carriers are located at the interface, *i.e.*, the nominal conduction channel.[22,23] Similarly, in LTO/STO SLs, the LDHM carriers spread into the STO, while the HDLM carriers are mainly located near the interface.[12,13] In the fractional SLs, another interesting behavior could be observed: LDHM carriers are almost independent on $x$, while HDLM carriers show large $x$-dependence. This suggests that the small LDHM carriers that are spread into STO are not strongly influenced by the amount of La in the 2DEG (as long as $x \geq 0.50$), while the transport properties of HDLM carriers that are located at the fractionally $\delta$-doped layer are mostly governed by $x$ within the nominal conduction channel. The transport properties are further investigated by a theoretical study. Figure 4a summarizes the spatial dependence of $n$ for LSTO/STO SLs extracted from density functional theory (DFT) calculations. It correctly predicts the $x$-dependence of $n_1$ near the LSTO $\delta$-doped layer. This implies that fractional $\delta$-doping can effectively tune the properties of HDLM carriers.

In this aspect, it would be quite intriguing to compare the HDLM carriers in 2DEG of LSTO with the carriers in 3D bulk LSTO. This comparison may provide a general idea on the conduction mechanism of filling-controlled oxide 2DEGs. According to a previous report, $n$ increases linearly with $x$ from $x = 0$ to 0.95 in bulk LSTO.[15] When $x$ reaches 1, the sample becomes an insulator. The increase of $n$ up to $x =$ 0.95 in bulk LSTO follows the ideal calculation based on the assumption that each substituted La site supplies one electron to the system, as shown in the black line in Figure 3c. More importantly, an increase in $m^*$ with $x$ has also been observed by various measurements within the Fermi liquid model, due to the strong electron correlation.[15,16]

In LSTO 2DEG, $n_1$ increased with $x$ as observed in bulk. Quantitatively, however, $n_1$ was only about half compared to that for the bulk. For LTO/STO integer SLs, the $d$ electrons in LTO can spill over into STO.[6,7] In this case, only about half of the induced electrons are mobile within the channel of LSTO. Therefore, a systematic decrease in $n_1$ with decreasing $x$ can be attributed to the bulk-like effect in combination with the electronic reconstruction in the fractional SLs.

The consistent behavior of $n$ between LSTO 2DEG and 3D bulk suggests that $m^*$ might also be increased with $x$ in 2DEG. To examine this idea, we performed DFT calculations, since it was rather difficult to experimentally determine $m^*$ of the SLs. Figure 4b shows a band diagram of the LSTO/STO SLs. Note that since $n_1$ is at least an order of magnitude larger compared to $n_2$, we should mostly see the response of the HDLM carriers. For $x = 0.25$ SL, only one band was barely crossing the Fermi level ($E_F$), indicating a semiconducting ground state, consistent with the experiment. For $x = 0.50$ SL, the bottom of the lowest-lying two conduction bands were filled, indicating a metallic state. When $x$ increased further, four more bands crossed $E_F$ and contributed to the metallic conduction, as more clearly shown in the magnified insets of Figure 4b. Using the curvature of the relevant electronic bands crossing $E_F$, we calculated the $m^*/m_e$ as summarized in Table I. In all cases, we found that the lowest energy bands had $m^*$ of ~0.5 $m^*/m_e$ , consistent with previous theory and experiments on surface 2DEGs.[24,25] Note that the sixth band was rather heavy compared to the low-lying bands. As $x \geq 0.75$, this heavy band started to contribute to the conduction, which should increase the overall $m^*$. For $x = 0.50$ SL, since only two light bands were involved, $m^*$ was rather low, resulting in an enhanced $\mu_1$. In addition, the band diagram suggested that fewer numbers of bands were actually contributing for $x = 0.50$ SL, and a decrease in the inter-subband scattering could be considered.[26]

It is noteworthy that $n$ in $x = 0.50$ SL is within the regime of semiconductor 2DEGs (typically about ~$10^{13}$ $cm^{-2}$). Moreover, the decreased $m^*$ also approaches the values observed for semiconductors. This suggests that by artificially controlling the complex band filling in oxide 2DEG systems, one could optimize the transport properties to achieve less interacting electron gas as in conventional semiconductors. Moreover, while band gap engineering and modulation doping are often used to fabricate semiconductor-based 2DEGs, we show that oxide-based 2DEGs can further utilize simple doping mechanisms to systematically control $n$ and $m^*$, adding another degree of freedom for manipulating transport properties and realizing high performance oxide electronics. Finally, the drastic increase in $\mu_1$ for $x = 0.50$ SL is indeed a significant breakthrough in oxide based 2DEG in the aspect of device application, since the device performance is mostly determined by the high-density carriers, which dominate the global conduction.

In summary, by studying fractional-layer control of electronically reconstructed interfaces' transport properties in artificial oxide superlattices, we showed that the characteristics of mobile charge carriers including carrier density, effective mass, and mobility could be efficiently tuned. Such control of charge carriers enables us to induce an insulator-metal transition even in a unit-cell thin, delta-doped 2D oxide system, adding another degree of freedom in designing oxide-based heterostructures and electronic devices.

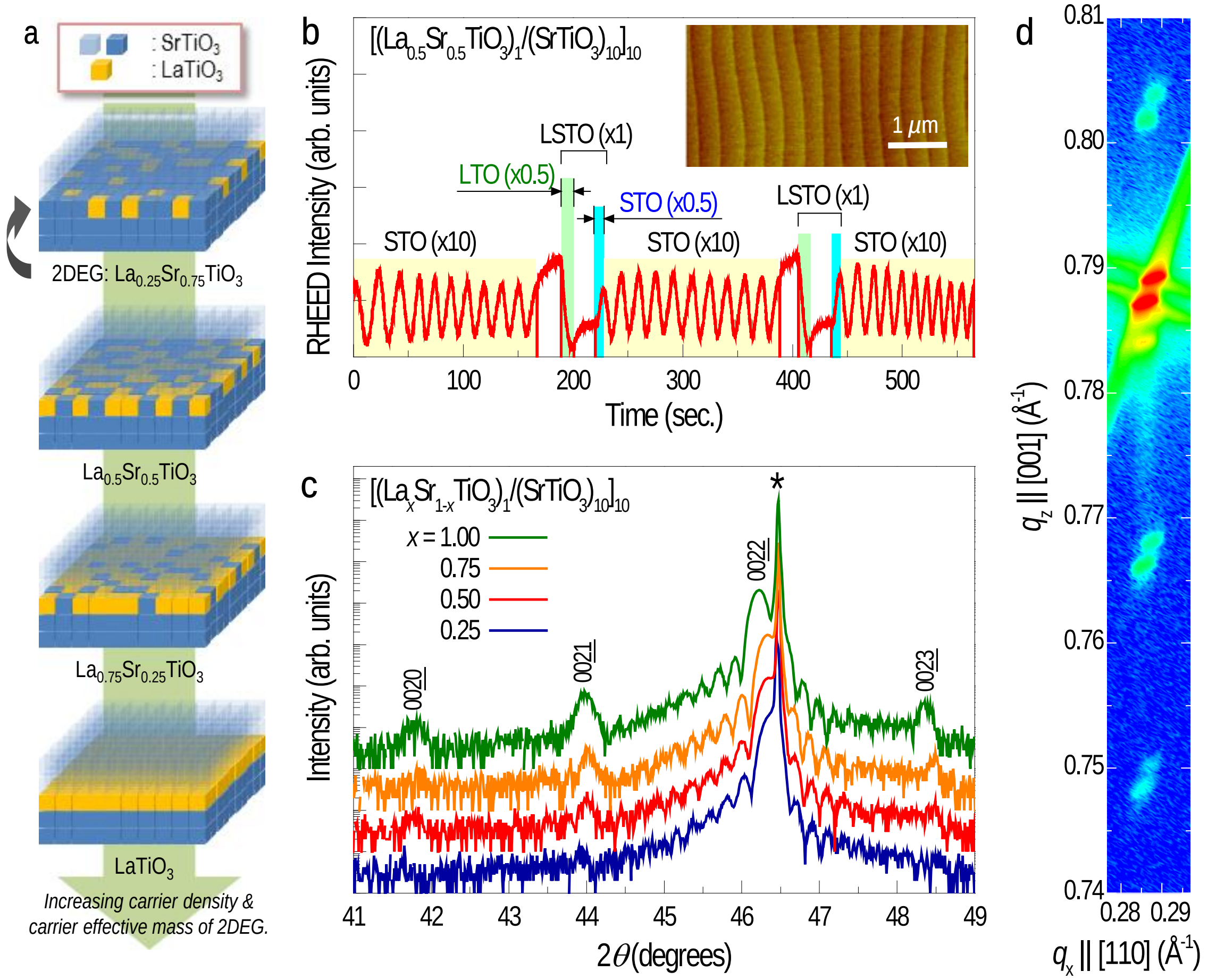

**Figure 1.** Artificial 2D crystals. (a) Schematic diagram of fractionally $\delta$-doped layers and controlling $n$ and $m^*$ of oxide 2DEGs. STO and LTO building blocks (perovskite u.c.) are shown in blue and yellow cubes, respectively. (b) RHEED intensity oscillations during the growth of a $La_{0.5}Sr_{0.5}TiO_3$/STO SL. Bright yellow-, green-, and blue-shaded regions represent oscillations recorded while growing 10 u.c. of STO, 0.5 u.c. of LTO, and 0.5 u.c. of STO, respectively. The inset shows a topographic image by atomic force microscopy of the $La_{0.5}Sr_{0.5}TiO_3$/STO SL showing an atomically well-defined single u.c. step-terrace structure conserved even after growth. (c) XRD $\theta$-$2\theta$ scans of SLs around the substrate STO 002 peak (*), clearly revealing the SL satellite peaks and Kiessig fringes from the highly accurate growth control. The scans are vertically shifted for clarity. (d) X-ray reciprocal-space map of an $x$ = 1.00 SL around the substrate $SrTiO_3$ 114 Bragg peak.

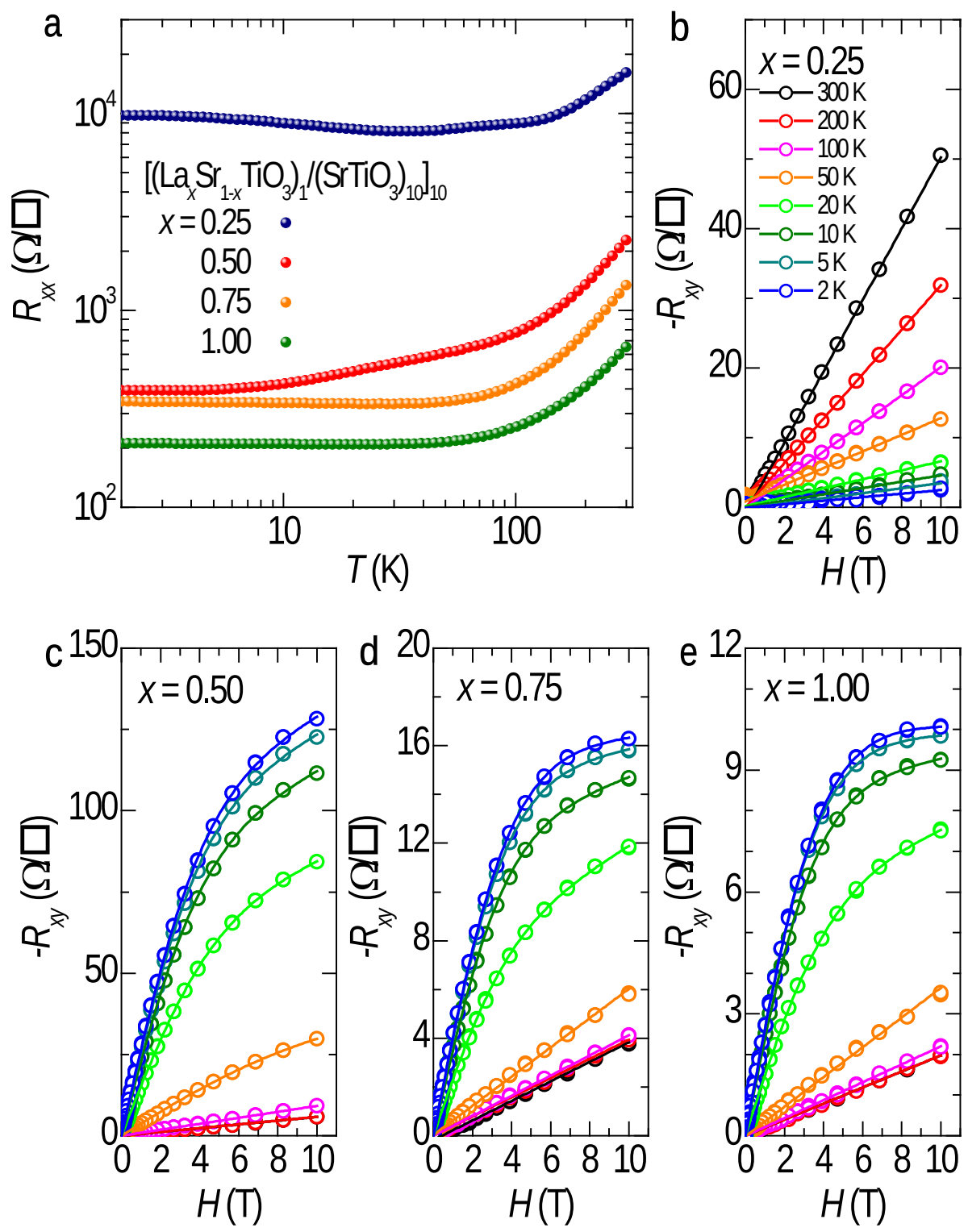

**Figure 2.** (a) Sheet resistance $R_{xx}(T)$ of LSTO/STO SLs and $R_{xy}(H)$ at various $T$ for (b) $x$ = 0.25, (c) $x$ = 0.50, (d) $x$ = 0.75, and (e) $x$ = 1.00 SLs. The experimental (symbols) and the fitting (lines) results are consistent with each other, validating the multichannel conduction model.

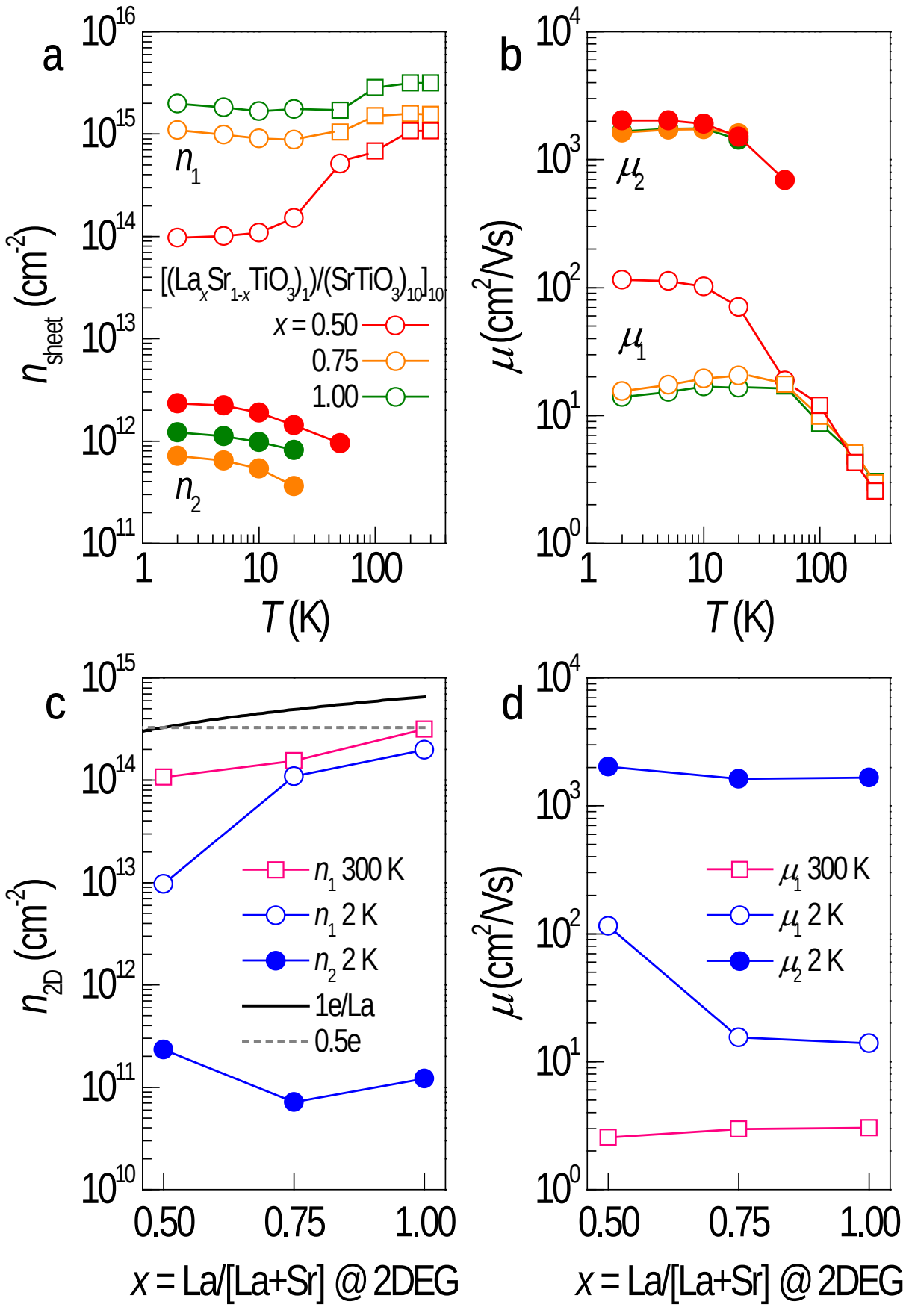

**Figure 3.** $T$-dependence of (a) $n_{sheet}$ and (b) $\mu$ of LSTO/STO SLs. Empty (filled) symbols indicate HDLM (LDHM) charge carriers. Data represented as circles (squares) are results from the two-channel model (linear) fit. $x$-dependence of (c) $n_{2D}$ and (d) $\mu$ of LSTO/STO SLs. $n_{2D}$ is $n_{sheet}$ of a single 2DEG, i.e. $n_{2D}$ = $n_{sheet}$/number of LSTO layers. The black solid line represents the expected $n$ assuming one electron is nominally supplied for each La substitution in the bulk LSTO system. The grey dashed line corresponds to 0.5 electrons per u.c.

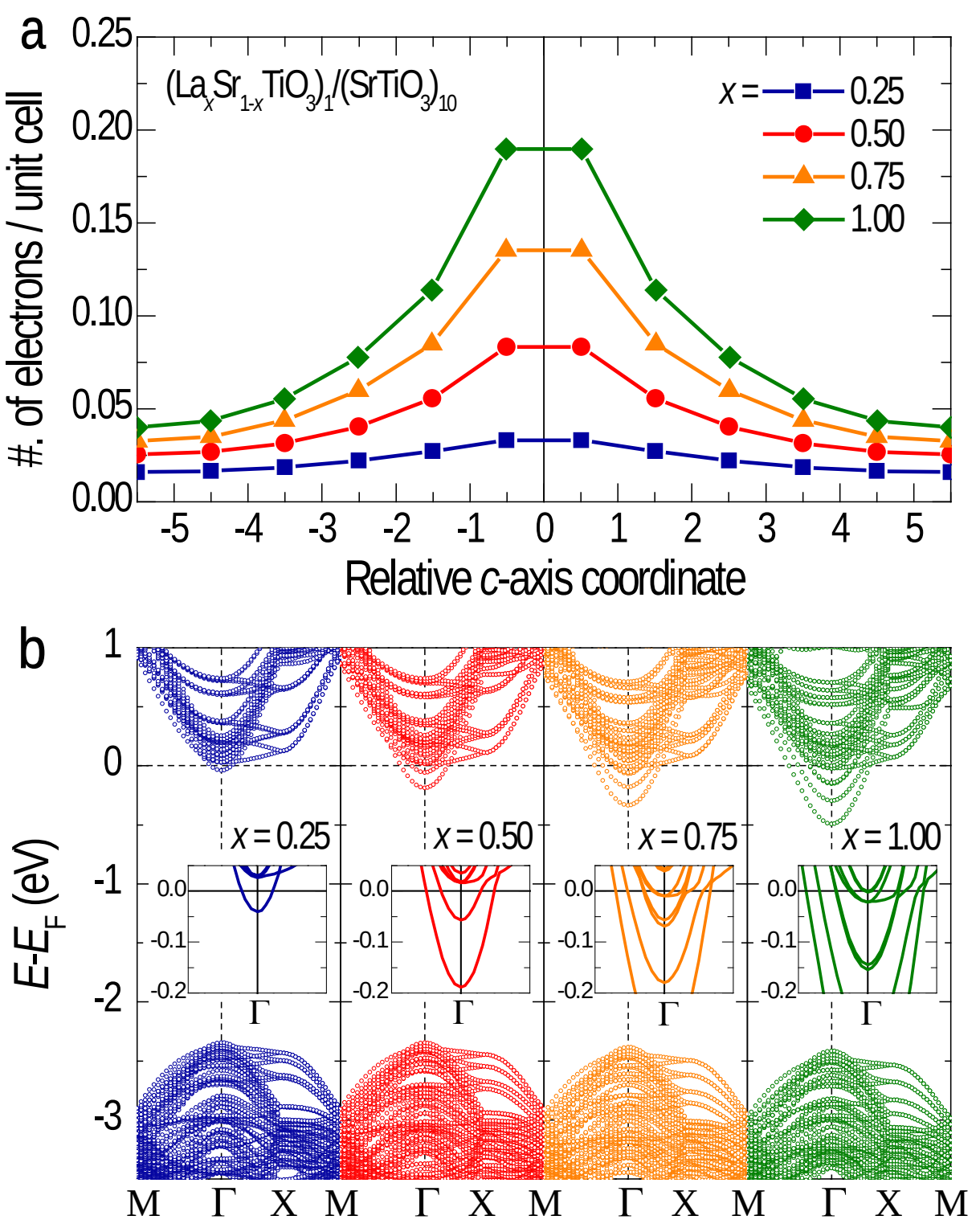

**Figure 4.** (a) Charge density profiles as a function of relative *c*-axis coordinate of Ti for LSTO/STO SLs. Data points are indicative of relative Ti coordinate (La ion is located at zero in *x*-axis). (b) Band structures of LSTO/STO SLs for $x$ = 0.25, 0.50, 0.75, and 1.00. The bottom of the conduction band is gradually filled as $x$ increases. The insets show the enlarged regions near $E_F$.

**Table 1.** Summary of relative effective masses, $m^*/m_e$, of the partially occupied bands for LSTO/STO SLs from DFT.

| $x$ | band | $m^*/m_e$ |
|---|---|---|
| 0.50 | 1 | 0.48 |
| | 2 | 0.54 |
| 0.75 | 1 | 0.49 |
| | 2 | 0.58 |
| | 3 | 0.52 |
| | 4 | 0.55 |
| | 5 | 0.69 |
| | 6 | 4.51 |
| 1.00 | 1 | 0.46 |
| | 2 | 0.57 |
| | 3 | 0.51 |
| | 4 | 0.52 |
| | 5 | 0.69 |
| | 6 | 4.84 |

Note that the curvature for low $m^*$ bands is symmetric around Γ. *i.e.* the $m^*$ along the Γ-X and Γ-M directions are essentially equal. High $m^*$ bands are computed along the Γ-X direction only.

The authors thank J. H. You, J. H. Lee, S. Okamoto, K.-S. Yi, and J. S. Kim for valuable discussions. This work was supported by the U.S. Department of Energy, Basic Energy Sciences, Materials Sciences and Engineering Division. This research used resources of the National Energy Research Scientific Computing Center, supported by the Office of Science, U.S. Department of Energy under Contract No. DEAC02-05CH11231. SL was supported by KIST grants, 2E22121 from Ministry of Educational Science and Technology.